\documentclass[aps,pra,10pt,twocolumn,amsmath,amssymb,
    longbibliography,superscriptaddress,nofootinbib]{revtex4-1}

\usepackage{graphicx}
\usepackage{amsmath}
\usepackage{dcolumn}
\usepackage{bm}
\usepackage{xcolor}
\usepackage{graphicx}
\usepackage{physics}
\usepackage{mathptmx}
\usepackage{lipsum}
\begin{document}

\title{Oxygen vacancies nucleate charged domain walls in ferroelectrics}

\author{Urko Petralanda}
 \affiliation{Computational Atomic-Scale Materials Design (CAMD), Department of Physics, Technical University of Denmark, 2800 Kgs. Lyngby, Denmark}
\author{Mads Kruse}%
 \affiliation{Computational Atomic-Scale Materials Design (CAMD), Department of Physics, Technical University of Denmark, 2800 Kgs. Lyngby, Denmark}
\author{Hugh Simons}%
 \affiliation{Department of Physics, Technical University of Denmark, 2800 Kgs. Lyngby, Denmark}
\author{Thomas Olsen}%
 \email{tolsen@fysik.dtu.dk}
  \affiliation{Computational Atomic-Scale Materials Design (CAMD), Department of Physics, Technical University of Denmark, 2800 Kgs. Lyngby, Denmark}

\date{\today}

\begin{abstract}

We study the influence of oxygen vacancies on the formation of charged 180$^\circ$ domain walls in ferroelectric BaTiO$_3$ using first principles calculations. We show that it is favorable for vacancies to assemble in crystallographic planes, and that such clustering is accompanied by the formation of a charged domain wall. The domain wall has negative bound charge, which compensates the nominal positive charge of the vacancies and leads to a vanishing density of free charge at the wall. This is in contrast to the positively charged domain walls, which are nearly completely compensated by free charge from the bulk. The results thus explain the experimentally observed difference in electronic conductivity of the two types of domain walls, as well as the generic prevalence of charged domain walls in ferroelectrics. Moreover, the explicit demonstration of vacancy driven domain wall formation implies that specific charged domain wall configurations may be realized by bottom-up design for use in domain wall based information processing.
\end{abstract}

\maketitle

Domain walls (DWs) in ferroelectric crystals are two-dimensional topological defects separating domains of distinct directions of the spontaneous polarization. They are ubiquitous, and significantly affect physical properties \cite{Catalan2005,Bednyakov2018,Nataf2020} such as polarization switching \cite{LiuRappe2016,LiPan2016}, dielectric permittivity \cite{Fousek1966,Zubko2016} and piezoeletric response \cite{Arlt1980,Rojac2015}. In addition, DWs are typically highly mobile \cite{Tybell2002} and their position may be controlled by external electric fields. The versatile properties of DWs \cite{YangRamesh2010, WuLai2017} have thus opened exciting avenues for applications in electronics, such as diodes \cite{Whyte2015} and non-volatile memory devices\cite{Sharma2017,Jiang2018}, and are promising candidates as building blocks for the next generation of photovoltaics \cite{Cook2017, Zenkevich2014}.

The properties of individual DWs strongly depend on the orientation of the spontaneous polarization ($\vb{P}$) with respect to the DW. When the normal component of $\vb{P}$ changes across the wall, a net bound charge is created in the DW \cite{Seidel2009}, giving rise to electric fields that typically far exceed the coercive field for polarization reorientation. Such a charged domain wall (CDW) would be highly unstable without a mechanism to screen the bound charge \cite{Ivanchik1973,Sluka2013,Selbach2018}. As such, unambiguous verification that CDWs are present in proper ferroelectric crystals \cite{Seidel2009, Gureev2011,Sluka2013,Sifuna2020} has lead to the conclusion that charged impurities must play a fundamental role in stabilizing CDWs \cite{Bednyakov2015}.

The role of oxygen vacancies (V$_{\mathrm{O}}$) as a stabilizing agent for CDWs has been studied both experimentally and by simulations \cite{HeVander2003,Gopalan2007,Shilo2004,Bismayer2005, GongLiu2018,GengMa2020,Borisevich2014}. In addition, it is known that V$_{\mathrm{O}}$s may assemble in perovskite lattice planes under certain conditions \cite{Scott2000,Kim2015}. However, these two effects have generally been regarded as unrelated, as the most widely accepted view is that V$_{\mathrm{O}}$s serve to stabilize CDWs that have formed spontaneously, or by other means. However, previous studies have also shown that V$_{\mathrm{O}}$s located at axial sites of TiO$_6$ octahedra in PbTiO$_3$ create a displacement of the Ti atom \cite{Park1998} and a corresponding dipole moment. This suggests an alternative view in which vacancies may in fact facilitate the formation of CDWs instead of simply delivering a stabilizing charge distribution. The notion that V$_{\mathrm{O}}$s may directly nucleate CDWs has broad implications for the understanding and application of CDWs in general. However, the formation mechanisms of CDWs remain elusive due to the apparent strong instability of CDWs and the technical challenges associated with carrying out \textit{ab initio} studies of CDWs in both pristine and doped ferroelectrics.

In this letter, we describe results from first-principles calculations of 180$^\circ$ domain walls in BaTiO$_3$ showing that it is favorable for oxygen vacancies to accumulate in planes, and that such accumulation will give rise to charged domain walls forming spontaneously. In particular, we demonstrate that the screening of the bound charge of head-to-head (HH) CDWs occurs through the filling of local conduction bands, regardless of vacancies. The negative charge at tail-to-tail (TT) CDWs is screened by the positive charge from the vacancies, thus quenching the $p$ type conductivity characterizing TT CDWs in the pristine system.
\begin{figure}
    \includegraphics[width=\linewidth]{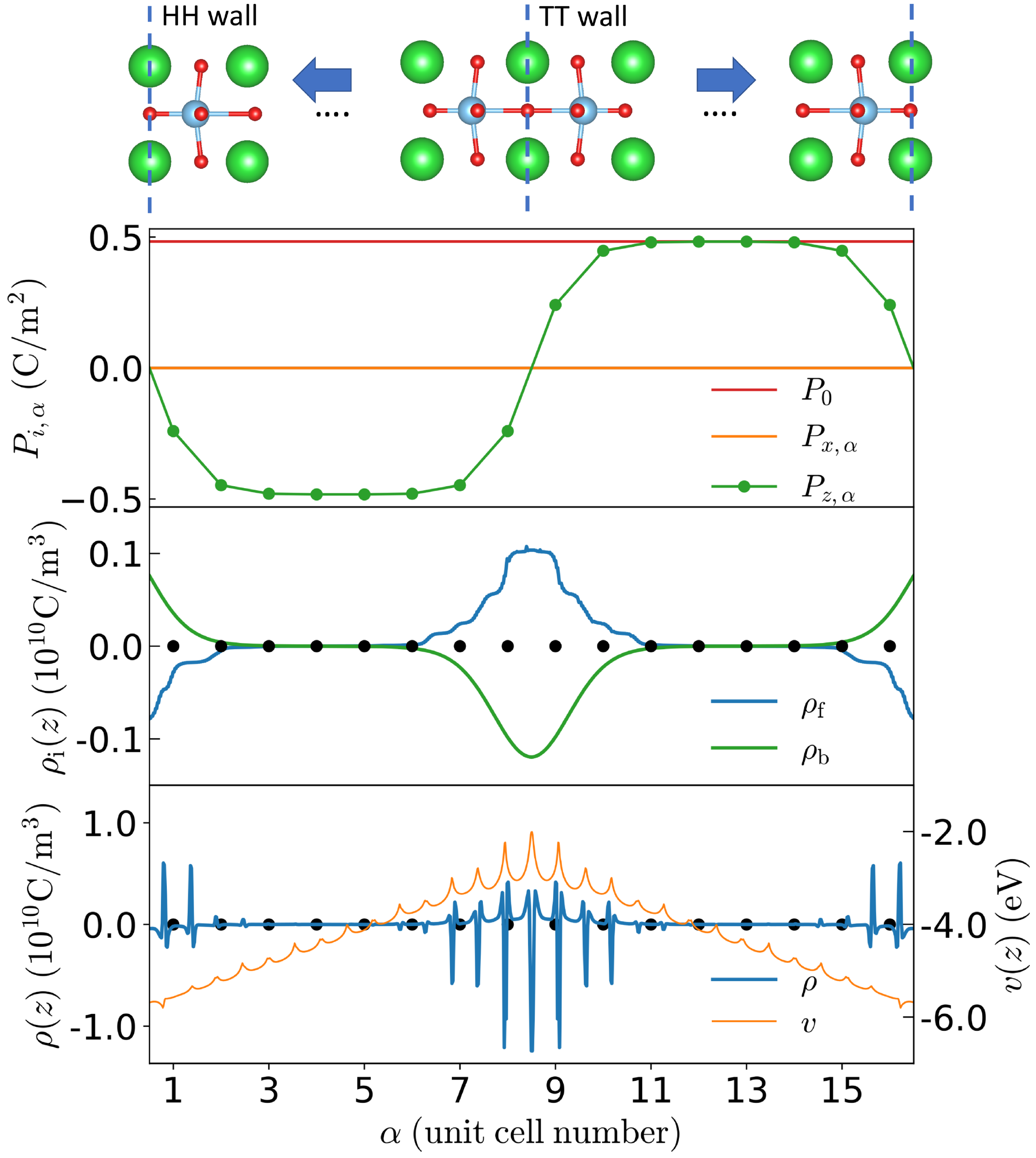}
    \caption{
        \label{fig:stru}
        Properties of the unrelaxed CDWs in a $1\times1\times16$ supercell. Top: schematic atomic structure with exagerated displacements. The blue arrows indicate the direction of polarization; green, blue and red spheres indicate Ba, Ti and O atoms, respectively. The dots indicate positions of Ti atoms in the individual unit cells. Second from top: polarization profile of the supercell. Second from bottom: bound $\rho_\mathrm{b}$ and free $\rho_\mathrm{f}$ charge density. Bottom: electrostatic potential energy  $v(z)$ averaged over the plane orthogonal to $z$ and total charge density $\rho$ obtained from a sliding window average.}
\end{figure}

BaTiO$_3$ is a perovskite oxide and a prototypical proper ferroelectric. It undergoes a well known phase transition sequence: from a cubic (Pm$\bar{3}$m) to tetragonal (P4mm) structure at $\sim$404 K, where a polarization emerges in the (001) direction; a subsequent first-order transition to an orthorhombic (Amm2) structure at $\sim$273 K where the polarization rotates to the (011) direction, and finally another first-order transition at $\sim$183 K to a rhombohedral (R3m) structure with polarization in the (111) direction \cite{Ravel1998}. Due to its simplicity and importance in applications, BaTiO$_3$ has served as a generic platform for studying neutral and charged DWs in the past \cite{Hlinka2006,Sluka2013}. Here we will focus on the tetragonal phase, but expect that the conclusions will hold true for other phases and similar compounds.

Our calculations were performed in the framework of Density Functional Theory (DFT), implemented in the GPAW electronic structure package \cite{Enkovaara_2010, Hjorth_Larsen_2017} using the projector-augmented wave method \cite{blochl}, the  Perdew-Burke-Ernzerhof exchange correlation functional \cite{Perdew1996}, and a plane wave basis. We used a plane-wave cutoff of 700 eV and a $\Gamma$-centered Monkhorst-Pack $k$-point grid with a density of 6 $\mathrm{\AA}^{-1}$. The parameters were converged with respect to the Born effective charges in the bulk structure. Forces were relaxed below 0.01 eV/{\AA} in all cases and below 0.005 eV/{\AA} for structures demanding better accuracy (for further details on the atomic structure and Born effective charges we refer to the Supplemental Material (SM)\cite{Supp2020}.

We begin by considering 180$^\circ$ CDWs in BaTiO$_3$ without any vacancies. This will serve as a reference system that allows us to gain insight into the influence of vacancies on the electronic properties of CDWs. In addition, it is known that a dilute distribution of CDWs in oxide perovskites can be stable and robust in defect-free thin films \cite{Gureev2011}. We thus construct a $1\times1\times16$ supercell of tetragonal BaTiO$_3$ and divide it into two areas of opposite polarization parallel to the long axis of the supercell. The polarization of the subcell $\alpha$ can be written as  $P_\alpha=\sum_{j,a\in\alpha}Z_{ij}^{*a}d_j^a$ \cite{Meyer2002}, where $d^a_j$ is the displacement of atom $a$ in direction $j$, $Z_{ij}^{*a}$ is the Born effective charge tensor of atom $a$ and the sum runs over atoms in unit cell $\alpha$. Since the Born effective charges depend on the local electronic structure, we choose $Z_{ij}^{*a}$ as the average of the values in the cubic and tetragonal phases of BaTiO$_3$ \cite{Supp2020}. The atomic displacements of the supercell are smoothed such that the polarization profile becomes $P_\alpha=P_0\tanh(z_\alpha  / \delta)$, where $z_\alpha$ is the center position of unit cell $\alpha$ and $P_0$ is the magnitude of the bulk polarization, 0.47 C/$\mathrm{m}^2$. We set $\delta$=1.75 as the structural width of the wall and adopt BaO centered DWs, since the V$_{\mathrm{O}}$s used in latter simulations are more stable at these planes. We did not relax the structure, as it would typically be driven into a single domain under relaxation with DFT, unless certain symmetries are fine-tuned to prevent it. Finally, we emphasize that, although this structure represents a somewhat artificial representation of a CDW, its primary intention is to unravel the basic principles of screening in the system. 

Fig. \ref{fig:stru} shows a schematic representation of the supercell including the profile of polarization per unit cell as calculated following the procedure in Ref \onlinecite{Meyer2002}. We also show the bound charge density $\rho_\mathrm{b}$ arising from the polarization profile, as well as the electrostatic potential obtained from DFT. As expected, a positive (negative) bound charge density peak is located at the HH (TT) wall and is accompanied by a minimum (maximum) of the potential. The integrated bound charge density at each of the walls has magnitude $2P_0$ by construction and the potential energy difference ($\Delta V$) between the walls is roughly 2.5 eV. Since the potential between the walls is linear, the electric fields inside the two domains can be regarded as constant and the {\it total} charge density at the CDWs are then related to the potential energy difference by Gauss law as
\begin{equation}
    \sigma_\mathrm{tot}(d)=\frac{\epsilon_0\Delta V/e}{d},
    \label{eq:capacitor}
\end{equation}
where $d$ is the distance between the walls and $\epsilon_0$ is the vacuum permittivity. Inserting the values obtained from DFT yields a charge density of $\sigma_\mathrm{tot}= 0.0056 \; \mathrm{C/m^2}$. This is two orders of magnitude smaller than the bound charge indicating that the bound charge is almost fully compensated by free charge. 

The mechanism behind the screening can be envisioned by considering two CDWs in close proximity with bound charge densities $\pm 2P_0$. Without any screening mechanism, the electric field between the walls would be determined by the polarization $P_0$ only. However, if the distance between the walls is increased, the potential difference between the walls increases (due to the constant electric field) and the conduction(valence) bands are lowered(raised) at the HH(TT) until they are aligned \cite{Gureev2011,Sifuna2020,Blugel2014}. At this point, charge will be transferred between the walls to align the Fermi levels at the walls, and the potential energy difference will be pinned at the value of the band gap. Thus, when the difference between the walls $d$ is increased in Eq. \eqref{eq:capacitor}, $\Delta V$ will remain fixed while the electric field and charge density at the walls decrease. 

For CDWs at large separation, the charge density will thus be finite, albeit orders of magnitude smaller than the bound charge. For the present calculations, we obtain a PBE band gap for BaTiO$_3$ of 2.0 eV, which is in agreement with the potential difference of 2.5 eV.  Note that for such a ``pristine'' CDW structure, the total charge density at the walls is roughly determined by the band gap of the material and the distance between the walls, whenever $d$ exceeds the ``short-circuit distanc'' $d_\mathrm{sc}=\epsilon_0E_\mathrm{gap}/2eP_0$. Integrating the charge density profile through the CDW (obtained from a sliding window average - see SM \cite{Supp2020} for details) yields  $0.006 \; \mathrm{C/m^2}$, which is in agreement with the result obtained from Eq. \eqref{eq:capacitor}.

\begin{figure*}
    \includegraphics[width=0.85\linewidth]{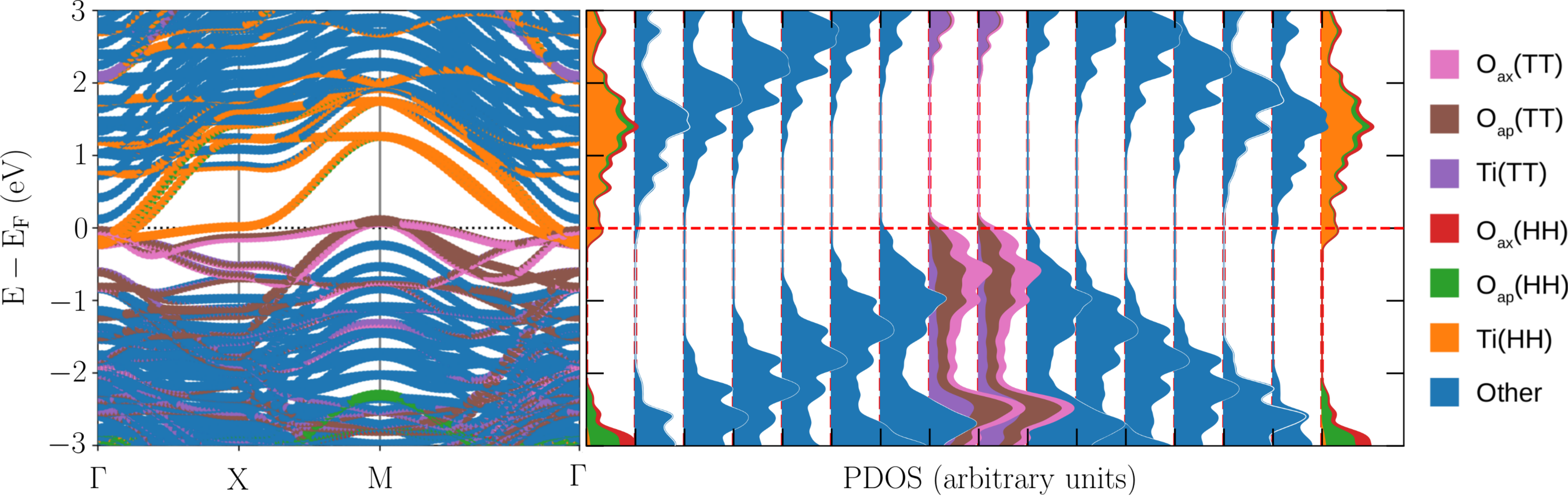}
    \caption{\label{fig:novac_band_wf}
        Left: projected band structure of the unrelaxed $1\times1\times16$ supercell with no vacancies. Right, the projected density of states resolved in individual unit cells.}
\end{figure*}
The screening described above is more clearly visualized from the band structure and projected density of states (PDOS) resolved in individual unit cells, which are shown in Fig. \ref{fig:novac_band_wf}. Due to the indirect band gap in BaTiO$_3$, the electron(hole) doping at the HH(TT) walls occurs at different locations in the 2D Brillouin zone. The band structure shows that the $\Gamma$-point mediates the electron doping at the HH wall, whereas the hole-doping mainly takes place at the $\mathrm{M}$ point. Resolving the PDOS in contributions from different unit cells yields a profile reminiscent of the electrostatic potential. For a given unit cell, the PDOS resembles that of bulk BaTiO$_3$, but shifted according to the local value of the electrostatic potential. The PDOS also implies that charge carriers in the vicinity of the Fermi level are strictly localized at the two CDWs as expected, which in turn implies that we can calculate the free charge density at the HH(TT) wall by adding the norm-squared wavefunctions of the conductive states below(above) the Fermi level (see SM for details \cite{Supp2020}). This procedure yields a free charge density of $\pm0.84 \; \mathrm{C/m^2}$ at the two walls, which (almost) cancel the bound charge density at the two walls as anticipated. The free charge density is shown in Fig. \ref{fig:stru} and exhibits a profile that compensates the bound charge.
 
\begin{figure}
    \includegraphics[width=\linewidth]{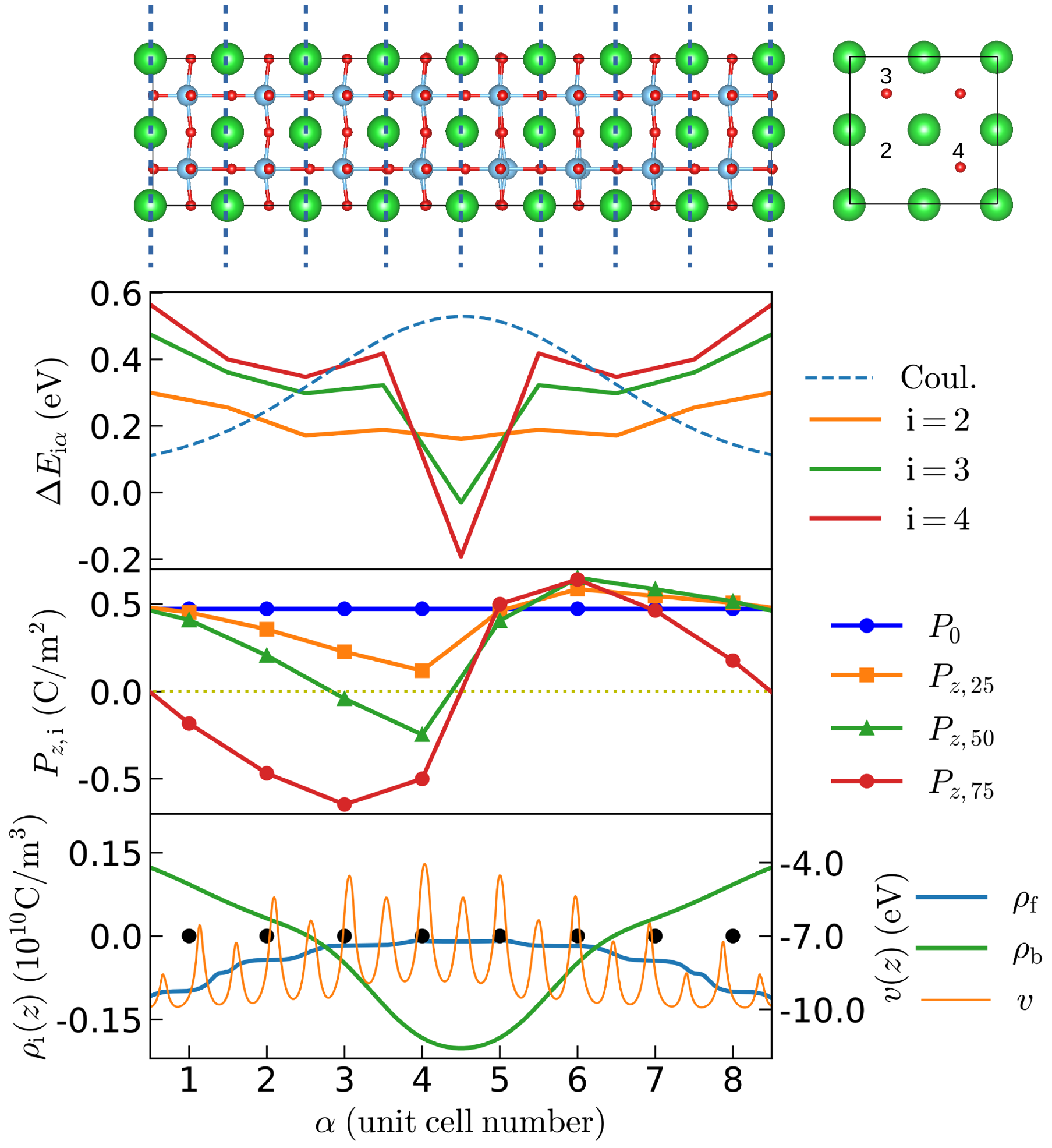}
    \caption{
        \label{fig:vacstru}
        At the top, the reference $2\times2\times8$ tetragonal BaTiO$_3$ supercell with one vacancy at a BaO plane. Below, the energy cost profiles for the $i^{th}$ vacancy at $\alpha$ position (see text). On the top-right, the section of the supercell corresponding to the central BaO plane where static vacancies are set. Second from the bottom, the polarization profiles of the relaxed supercell for $i=$ 0, 1, 2 and 3. At the bottom, bound and free charge density profiles for the relaxed $2\times2\times8$ supercell with two vacancies at the central BaO plane.}
\end{figure}

In order to study the role of V$_{\mathrm{O}}$s in the formation of CDWs, we use a bulk, monodomain $2\times2\times8$ supercell as the starting point. We find the most stable position for a planar distribution of oxygen vacancies is the BaO plane (see figure S2 \cite{Supp2020}), and as such we begin by introducing a single oxygen vacancy into a BaO plane, and fully relax the supercell (see the top part of figure \ref{fig:vacstru}). Following this relaxation, we see that the V$_{\mathrm{O}}$ strongly repel the neighboring Ti atom towards the opposite direction of the initial polarization, thereby decreasing the local polarization, as shown in Figure \ref{fig:vacstru}. This occurs through the breaking of the bonding orbital formed by the emptied O $p$ and the Ti $t_{2g}$, in a similar scenario to that described by Park \emph{et al} in Ref \cite{Park1998} for PbTiO$_3$.

We will now show that it is favorable for additional vacancies to migrate to the plane of the initial vacancy. To quantify the energy cost of adding additional V$_{\mathrm{O}}$s, we define the energy cost of adding the $i$'th vacancy in unit cell $\alpha$ relative to the energy cost of adding a single vacancy in a bulk BaO plane:
\begin{equation}
    \Delta E_{i\alpha}=E_{i\alpha}-E_{(i-1)\alpha_0}-(E_{1}-E_\mathrm{bulk}),\qquad i>1
    \label{eq:vacancy}
\end{equation}
where $E_\mathrm{bulk}$ is the energy of a single domain without vacancies, $E_1$ is the energy of the supercell with a single vacancy in any BaO plane and $E_{i\alpha}$ is the energy of the relaxed supercell with the $i$'th vacancy placed in unit cell $\alpha$ and $i-1$ vacancies at their optimal positions ($\alpha_0$). 

We first calculate $E_{2\alpha}$, starting with a fully relaxed configuration with a single vacancy in the BaO plane, and relax the structure with a second vacancy in all the possible BaO planes. The result, shown in Fig. \ref{fig:vacstru}, shows that the optimal position is the plane where the first vacancy was placed. We then calculate $E_{3\alpha}$ by the same procedure starting with the relaxed structure with two vacancies in a single BaO plane. Again, the optimal position is the BaO plane where the two other vacancies are located. The energy difference \eqref{eq:vacancy} at this optimal position is close to zero, implying that this position is as favorable as a BaO plane in a single domain without electrostatic potentials from a CDW. Repeating the same procedure with the fourth vacancy, we see that it experiences an even stronger attraction to the three initial vacancies (see Fig. \eqref{fig:vacstru}). 

This attraction between vacancies may seem counter intuitive, since they are expected to have a like charge of +2 and should thus repel one another. However, the vacancies induce a polarization that initiates a negatively charged DW, which in turn provides the attractive force between the vacancies. This process is evident in Fig. \ref{fig:vacstru}, where we show the polarization profiles for 1,2, and 3 vacancies at their optimal positions in the third panel, starting from the top. The first vacancy introduces a local polarization reminiscent of a negatively charged domain wall, which provides a potential well for the second vacancy. If the second vacancy is placed at the optimal position at the wall the local polarization is distorted further, giving rise to an even deeper potential well for the third vacancy and so forth. 

The energetics of the vacancies described above suggests that accumulation of oxygen vacancies in planes occurs naturally in the material, and that such accumulation is accompanied by the formation of a negatively charged domain wall. The first vacancy introduces a weak polarization that gives rise to a potential well, and any other vacancy that sees this potential (which is long ranged) will be attracted to the BaO plane where the initial vacancy is situated (see Fig, \ref{fig:vacstru}).

Our simulations use a $2\times2\times8$ supercell such that a single vacancy corresponds to a 25 \% oxygen depletion in a single BaO plane. As such, it is then natural to ask whether the effect of oxygen accumulation will hold true at realistic vacancy distributions. In this sense, for a random distribution of vacancies there will inevitably be some planes with a higher vacancy concentration than others, and such planes will act as sinks for vacancies of the neighboring environment (though the driving force for accumulation may be smaller than that calculated here). The actual accumulation will be mediated through thermally assisted diffusion, however, a quantitative estimation of diffusion rates is beyond the scope of this work.

It is instructive to compare the charge density profiles shown in Fig. \ref{fig:vacstru}, with the case without vacancies shown in Fig. \ref{fig:stru}. We see a similar picture at the HH wall where the negative free charge screens the positive bound charge, although the extent of the wall is somewhat larger in the present case. In contrast, at the TT wall there is nearly no free charge since the bound charge already compensates the positive charge of the V$_{\mathrm{O}}$s. The small concentration of free charge at the TT wall explains the suppression of electric conductivity at TT walls in the presence of V$_{\mathrm{O}}$s observed in BaTiO$_3$ \cite{Bednyakov2015}. The conductivity mechanism of the HH wall remains essentially unaffected; Ti $t_{2g}$ states from the conduction band shift below the Fermi level in the HH wall area, similarly to the vacancy free case (see Figs. S1 and S3 \cite{Supp2020}). In contrast, the conductivity in the TT wall is almost completely quenched (see Fig S4 \cite{Supp2020}), due to the depletion of the O orbitals at the wall, which is in accordance with the absence of free charge in Fig. \ref{fig:vacstru}. The electrostatic potential profile between walls thus becomes flattened, alleviating the Coulomb interaction energy between the walls and significantly increasing their stability.

In conclusion, we have shown that, upon insertion of a single oxygen vacancy into tetragonal BaTiO$_3$, it is favorable for other vacancies to migrate to the BaO plane defined by the first vacancy and that a TT domain wall is formed in the process. The driving force is the negative bound charge emerging at the TT wall. It naturally follows from our calculations that oxygen vacancies are attracted to TT DWs, which has already been suggested in the past \cite{Scott2000,Bednyakov2015}. However, the fact that oxygen vacancies play a critical role in the {\it formation} of CDWs has not been demonstrated previously and provides a significant indication as to why and how CDWs form in the first place. Moreover, the implications of this mechanism are potentially far reaching, since it implies that a particular CDW distributions may be accomplished by simply seeding a ferroelectric with a suitable distribution of vacancies. Such control will be a crucial ingredient for the future development of domain wall nanoelectronics.

T.O. and U.P. were supported by the Villum foundation, grant number 00028145.

\providecommand{\noopsort}[1]{}\providecommand{\singleletter}[1]{#1}%

\clearpage
\newpage

\widetext
\begin{center}
\textbf{\large Supplemental Materials for: Oxygen vacancies nucleate charged domain walls in ferroelectrics}
\end{center}

\setcounter{equation}{0}
\setcounter{figure}{0}
\setcounter{table}{0}
\setcounter{page}{1}
\makeatletter
\renewcommand{\theequation}{S\arabic{equation}}
\renewcommand{\thefigure}{S\arabic{figure}}
\renewcommand{\thesection}{S\arabic{section}}
\renewcommand{\thetable}{S\arabic{table}}
\renewcommand{\bibnumfmt}[1]{[S#1]}
\renewcommand{\citenumfont}[1]{S#1}

\section{Additional structural details}
As building blocks of our supercells we use a relaxed tetragonal BaTiO$_3$ unit cells with lattice parameters and the atomic positions indicated in Table \ref{tab:positions}. We calculate the Born effective charges in the structure above and in the cubic phase, where we use a structure with lattice parameter $a=4.036 \;\mathrm{\AA}$. The non-vanishing components of the Born effective charge tensors obtained are listed in Table \ref{tab:borncharges}, in close agreement with \cite{Ghosez1995}. Our effective Born effective charge tensor is obtained by averaging these two tensors.

\begin{table}[htb]
\begin{ruledtabular}
\caption{Relaxed BaTiO$_3$ tetragonal structure with space group P4mm (no. 99) lattice parameters $a=b=3.997$\;\AA\ and $a=4.212$\;\AA.}
\label{tab:positions}\begin{tabular}{ccccc}

Atom&Wyckoff p.&$x$&$y$&$z$\\
Ba&1a&0.000&0.000&0.021\\
Ti&1b&0.500&0.500&0.539\\
O&1b&0.500&0.500&-0.029\\
O&2c&0.500&0.000&0.492\\
\end{tabular}
\end{ruledtabular}
\end{table}
\begin{table}
\begin{ruledtabular}
\caption{Non-vanishing components of Born-effective charge tensors obtained for BaTiO$_3$ tetragonal and cubic structures. Parallel and perpendicular directions are defined taking as a reference the Ti-O band directions, as done in Ref \onlinecite{Ghosez1995}. For the O atoms at $2c$ positions in the tetragonal structure, we define the perpendicular (to the Ti-O bond) I and II directions as being, in addition, parallel and perpendicular to the tetragonal axis, respectively.}
\label{tab:borncharges}\begin{tabular}{ccccccccccc}
Tetragonal\\
&$xx$,$yy$&&&&$zz$\\
Z$^*_{\mathrm{Ba}}$&2.68&&&&2.87\\
Z$^*_{\mathrm{Ti}}$&6.70&&&&4.83\\
\hline
&&$\perp$&&&&&&&$\parallel$\\
Z$^*_{\mathrm{O}_1}$&&-1.90&&&&&&&-4.00\\
&I&&II&&&&\\
Z$^*_{\mathrm{O}_2}$&-1.85&&-2.08&&&&&&-5.40\\
\hline
\hline
Cubic\\
&$xx$,$yy$,$zz$&\\
Z$^*_{\mathrm{Ba}}$&2.73&\\
Z$^*_{\mathrm{Ti}}$&7.28&\\
\hline
&& $\perp$&&&&&&&$\parallel$ \\
Z$^*_{\mathrm{O}}$&&-2.12&&&&&&&-5.77\\
\end{tabular}
\end{ruledtabular}
\end{table}

\section{Free charge density calculation method}

In order to calculate the free charge density across the supercell long axis plotted in figure 1 of the main text we use the following expression:

\begin{equation}
\begin{split}
\rho_f(z)=\frac{1}{AN_k}\sum_{\vb{k}}\int dxdy
\left[  \sum_{n_{E_F}}^{n^*_{VBM}}|\psi_{n\vb{k}}(\vb{r})|^2(\Theta(z-\frac{L}{4})-\Theta(z-\frac{3L}{4}))  
+\sum_{n^*_{CBM}}^{n_{E_F}}|\psi_{n\vb{k}}(\vb{r})|^2(\Theta(z+\frac{L}{4})- \Theta(z-\frac{L}{4}))  \right]
\end{split}
\label{eq:free}
\end{equation}

where  $\psi_{n\vb{k}}$ is the wave function at the $n^{th}$ band at k point $\vb{k}$, $A$ is the supercell area parallel to the walls, $N_k$ is the number of k points, $L$ is the supercell length, $x$ and $y$ are the coordinates parallel to the walls, and $n_{E_F}$, $n^*_{vBM}$ and $n^*_{CBM}$ represent the Fermi level, valence band maximum in the range $L/4<z<3L/4$ and conduction band minimum in the range $-L/4<z<L/4$, respectively; $\Theta(x)$ is the Heaviside step function and
$L$ the supercell length in $z$ direction.

In order to ensure an accurate sampling of the Brillouin zone and an accurate estimate of the free charge at the walls, we calculated it applying eq. \ref{eq:free} on a high k point density (12 points \AA$^{-1}$) scf calculation.

\section{Total net charge at the walls}

To gain insight on the charge distribution at the DWs we calculate the charge density per unit length in the supercell, $n(z)$. Within the projector-augmented wave formalism the all electron density $n_e(z)$ can be obtained, and we can represent the nuclear charges as point charges. Then, after integrating the electronic density over the coordinates parallel to the domain walls we get:

\begin{equation}
\label{eq:density}
\rho(z)=-\rho\textbf{}_e(z)+\sum_i Z_i\delta(z-z_i)
\end{equation}

where $n_e(z)$ is the all-electron density, $z_i$ are the coordinates of the nuclei and $Z_i$ are the nuclear charges. The charge density is rapidly varying at the atomic scale due to the localized nuclei and core electrons. Any integral of Eq. \ref{eq:density} will thus depend strongly on the domain of integration and the charge residing at the domain walls cannot be obtained in any sensible way. To remedy this to some extent we consider the sliding window average for the density and in addition we convolute the nuclear charge with a gaussian function, getting, for the net charge density as a function of the long axis coordinate $z$:

\begin{equation}
\label{eq:density}
\tilde{\rho}(z)=\frac{1}{c}\int_{z-c/2}^{z+c/2}\left(\rho_e(z')+ \sum_j \delta(z'-\xi_j) \sum_iZ_i g(\xi_j-z'_i) \right)dz'
\end{equation}

where we define the gaussian distribution as $g(x)=\frac{1}{\sigma\sqrt{2\pi}}\exp{\frac{-x^2}{2\sigma^2}}$ with an arbitrary $\sigma=20$ and $c$ is the unit cell length. This way we obtain the smoothened total charge density shown in Figure 1 in the main text. 

At the bottom part of figure 1 in the main text we see wide regions on both sides of the walls where the charge density is constant. This allows the computation of the net charge within the wall regions. 
We find a net charge density of -0.006\;$\mathrm{C/m^2}$ at the TT wall area and the exact opposite, 0.006\;$\mathrm{C/m^2}$ at the HH area.

Note that, if the supercell were insulating we would see, at any region enclosing one domain wall, an absolute net charge density of 2$P_0$ $\approx 0.94\;\mathrm{C/m^2}$, 
where 
$P_0$ is the absolute value of the spontaneous polarization of the bulk crystal ($P_0$ =0.47 C/m$^2$ for the tetragonal unit cell in our calculations).
This indicates near full charge compensation of the bound charge by free carriers. In addition, the bound charge enclosed at the walls is smaller than the one expected from the polarization change, given that the absolute value of the free charge calculated at the walls is 0.84\;$\mathrm{C/m^2}$. This is expected from the fact that the definition of Born effective charges becomes problematic in metal-ferroelectric interfaces \cite{Stengel2011}. It is instructuve to transform the charge densities at the walls into atomic units. We have for the net charge at the wall, for a=b $\approx 4 \mathrm{\AA}$ square unit cell parameter (area\;=\;$56.85\;a_0^2$), and taking into account that $1\;e=1.602 \times10^{-16}\;\mathrm{C}$: $\sigma_{tot}$=$\frac{0.006}{56.85}ea_0^{-2}$, which means that per unit cell area, the CDWs present a net charge of about 1/160$\;e$.

Finally, we also apply an analogous sliding window average to the electrostatic potential to obtain the smooth function $\tilde{v}$(z)

\begin{equation}
\label{eq:density}
\tilde{v}(z)=\frac{1}{c}\int_{z-c/2}^{z+c/2}v(z')dz'
\end{equation}

where $v$ is the electrostatic potential as defined in GPAW \cite{Enkovaara_2010_2}.

\section{Application of the pseudo-capacitor model}

In order to apply the equation 1 in the main text on the present system, we calculate the electrostatic potential originated by a net charge density of 0.006\;$\mathrm{C/m^2}$ (that we obtain at the walls by integrating the DFT all-electron density at the two walls). We have,

\begin{equation}
\label{eq:density}
\Delta V = \frac{\sigma_{tot}d}{\epsilon_0}
\end{equation}

In our system we have, in atomic units: $\sigma_{tot}$=$\frac{0.006}{56.85}ea_0^{-2}$; $d$=63.59 $a_0$; $\epsilon_0=\frac{1}{4\pi}e^2a_0^{-1}E_h^{-1}$.

And hence we get: $\Delta V\approx$0.084 $E_he^{-1}$, so $e\Delta V\approx$2.3 eV, which is essentially matching the electrostatic energy difference between the walls in figure 1 of the main text if we neglect the small oscillations at the vicinity of atoms.

\section{PDOS of 1x1x16 supercell without vacancies}
\begin{figure}
    \includegraphics[width=0.35\linewidth]{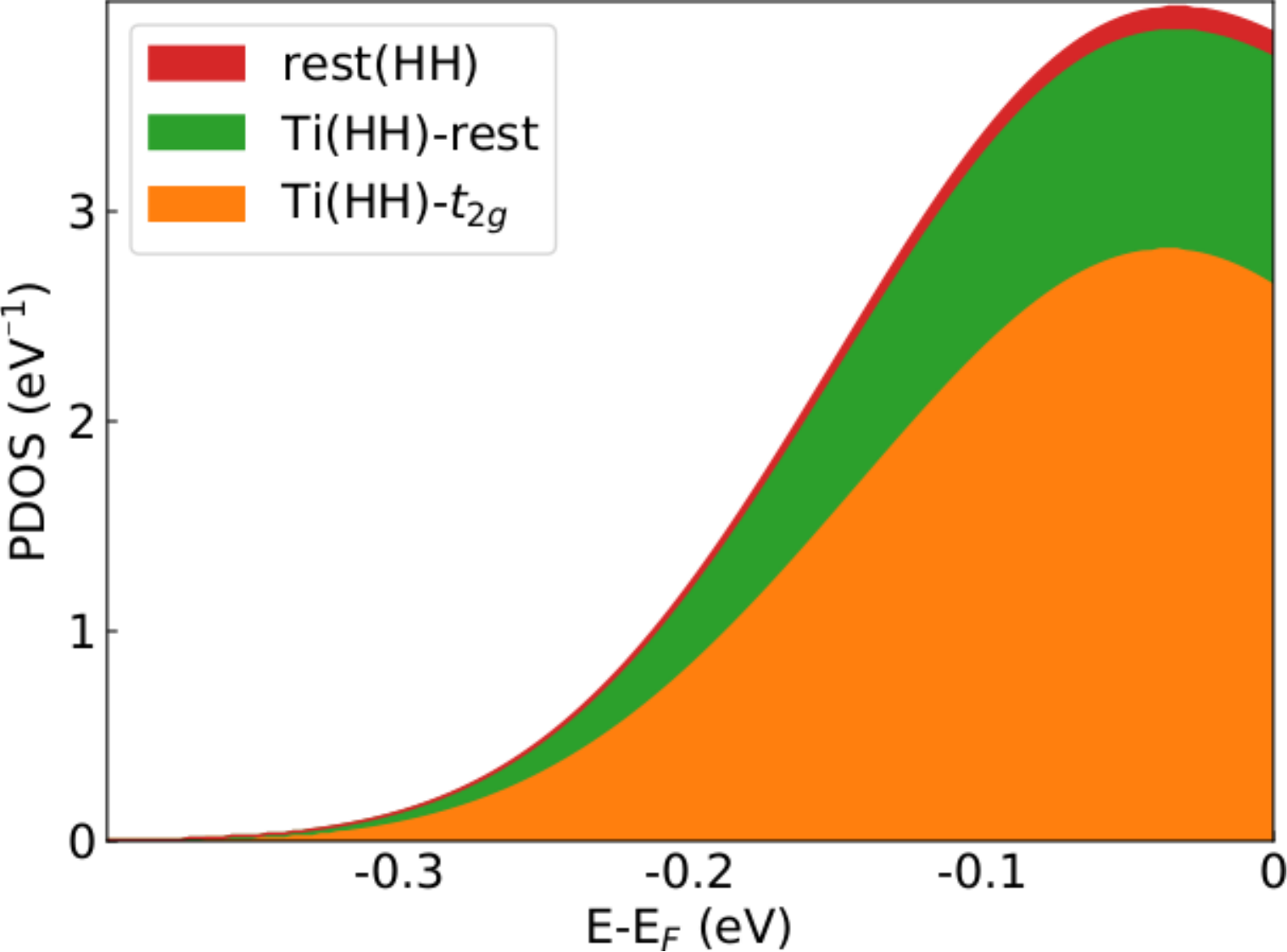}
    \caption{
        \label{fig:pdos}
        PDOS at the HH wall of the 1x1x16 supercell containing a HH and TT wall. Ti $t_{2g}$ states, Ti non-$t_{2g}$ states and the rest are differentiated with different colors. The Fermi level is set as the origin for the energies.
        }
\end{figure}

In figure \ref{fig:pdos} we show the projected density of electronic states (PDOS) on the HH wall in the 1x1x16 system with no vacancies and two charged domain walls. We see that $t_{2g}$ $d$ orbitals account for the majority of electronic states just below the Fermi level and therefore host the conduction band electrons that account for the conductivity of the HH wall.

\section{Determination of crystallographic plane for the vacancies}

\begin{figure}
    \includegraphics[width=0.75\linewidth]{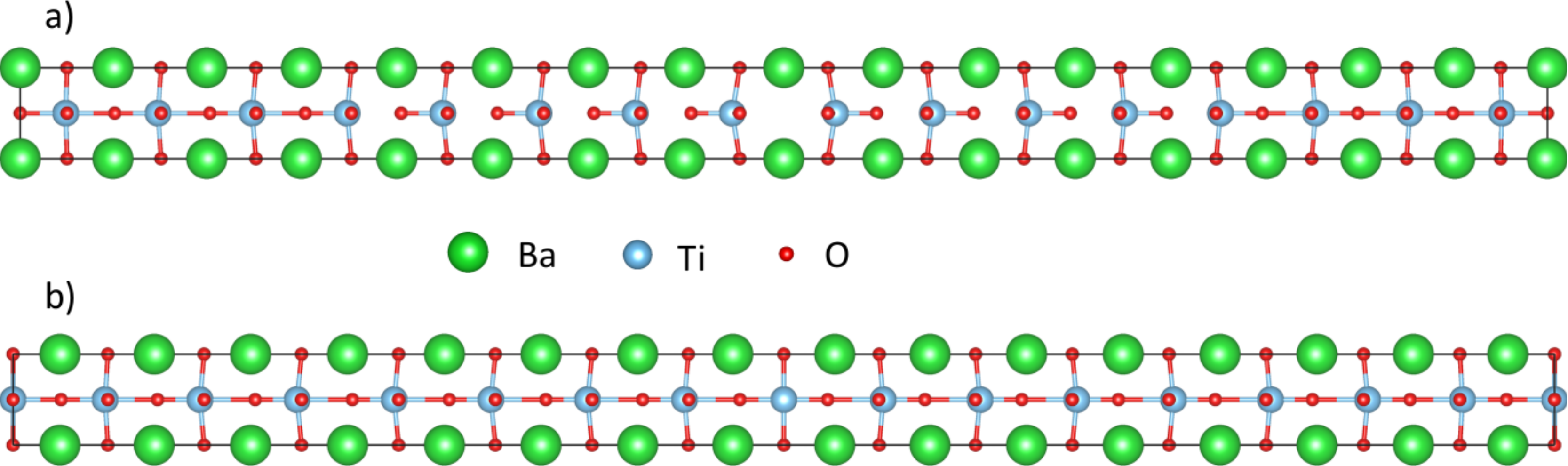}
    \caption{
        \label{fig:pzconv}
       The two DW configurations in 1x1x16 supercells used in order to determine the crystallographic plane where the oxygen vacancies are placed in the calculations. In a) BaO centered TT wall with vacancies, in b) TiO$_2$ centered TT wall with vacancies. The forces of both structures have been relaxed.
        }
\end{figure}

We take the 1x1x16 supercell containing a HH and TT wall in described in the main text and
discern between two cases, one with BaO centered walls and another one with TiO$_2$ centered walls. Then we empty one oxygen from the TT wall in each supercell and relax the forces. Both relaxed configurations are shown in figure \ref{fig:pzconv}; in \ref{fig:pzconv}.a the Ba centered case and in \ref{fig:pzconv}.b the Ti centered case. For the Ba centered case, there are no oxygen atoms left at the TT wall, while in the Ti centered case half of the oxygens are present after setting the vacancies. We find that the BaO centered walls with vacancies are more stable by about 0.6 eV. This occurs because the total depletion of oxygens at the TT wall allows for a stronger polarization change at the wall. Hence we select a BaO plane in the bulk supercell for the position of the initial V$_{\mathrm{O}}$ in our stability calculations. In fact, the TT wall grows naturally from an initial bulk configuration where an oxygen vacancy plane has been placed in a BaO plane.

\section{Electronic properties of relaxed supercells with oxygen vacancies}

\begin{figure}
    \includegraphics[width=0.35\linewidth]{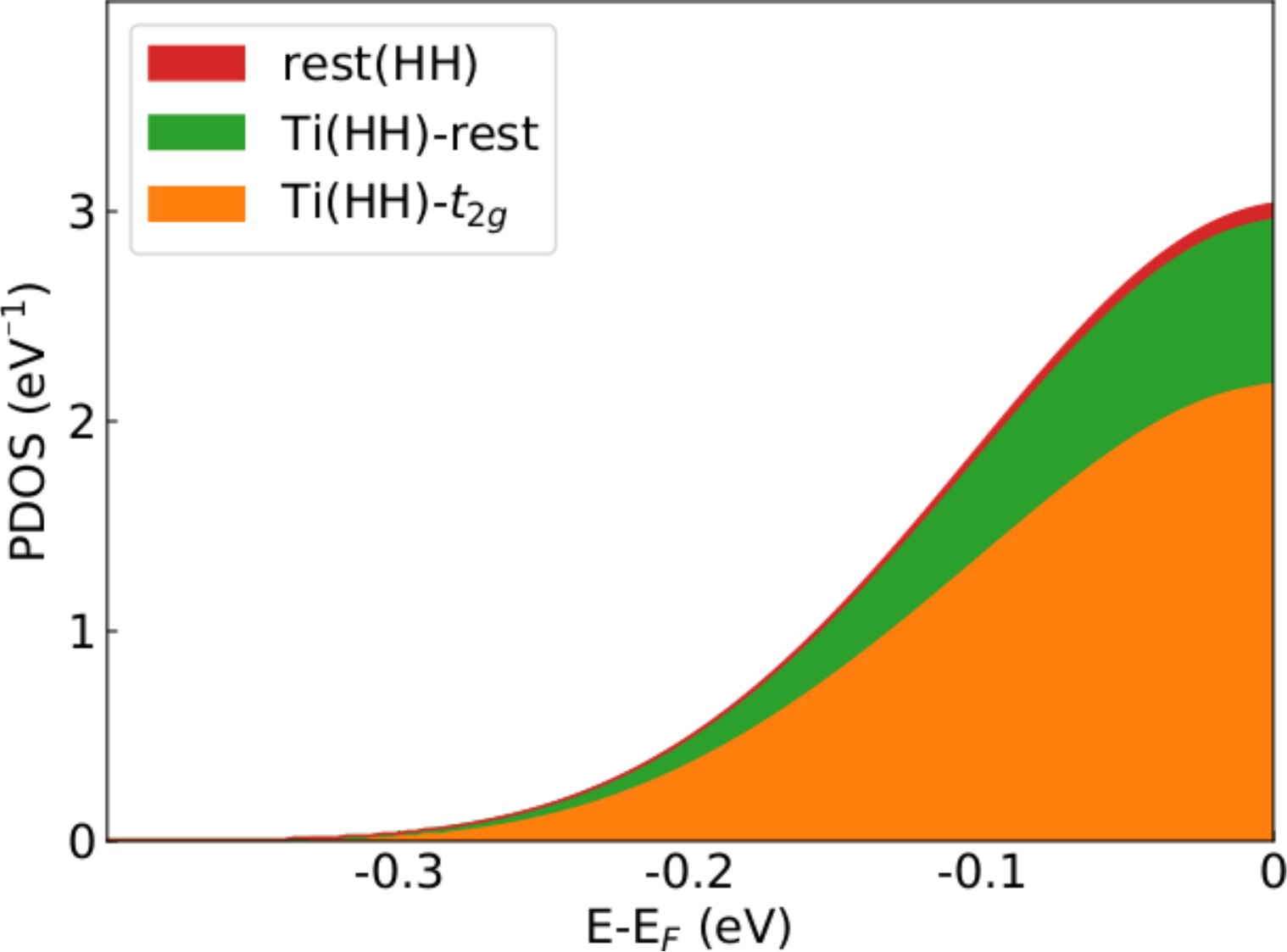}
    \caption{
        \label{fig:pdosvac}
        PDOS at the HH wall of the relaxed 1x1x16 supercell containing a HH and TT wall and an oxygen vacancy at the latter. Ti $t_{2g}$ states, Ti non-$t_{2g}$ states and the rest are differentiated with different colors. The Fermi level is set as the origin for the energies.
        }
\end{figure}

To show the little influence of the V$_{\mathrm{O}}$ on the electronic properties of the HH wall, we work on the longest supercell possible computationally in order to minimize the wall-wall interaction.
We take the relaxed 1x1x16 supercell containing a HH and TT wall in described in the previous section, with the oxygens in the BaO plane of the TT wall removed. We calculate the PDOS at the HH wall, in a similar way as done in the prevous section with the vacancy-free system. As shown in figure \ref{fig:pdosvac} we find that, again, Ti $t_{2g}$ states dominate this region of the DOS, suggesting little influence of the V$_{\mathrm{O}}$ on the ($n$ type) conductivity mechanism of the HH walls. 
\begin{figure}
    \includegraphics[width=0.55\linewidth]{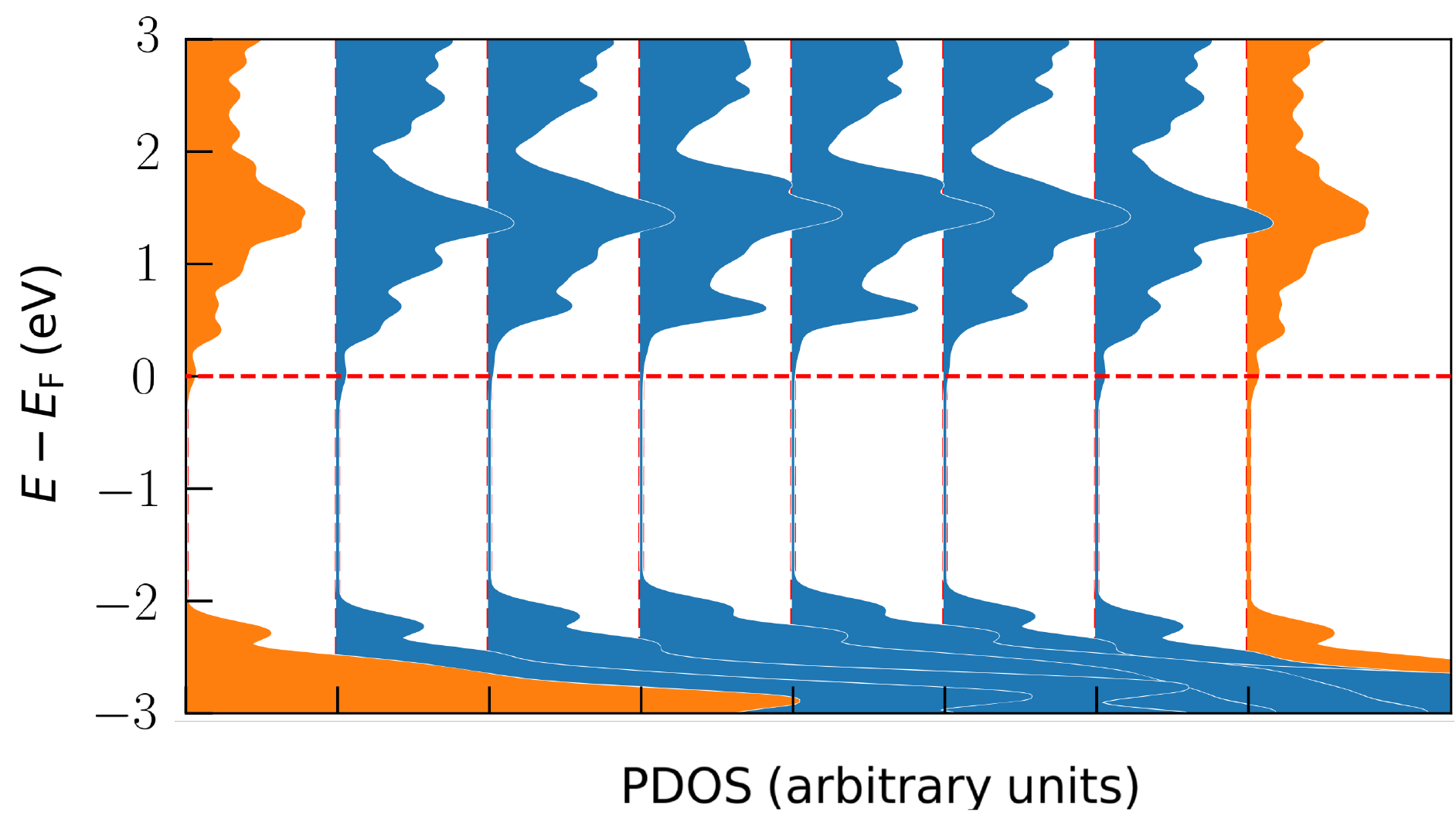}
    \caption{
        \label{fig:pdosall}
        Density of states of the relaxed 2x2x8 supercell with a HH and TT wall, with two vacancies at the latter, projected into the atomic orbitals of TiO$_2$ planes. The Fermi level is set as the origin for the energies.
        }
\end{figure}

The supercell above though, may not be appropriate to study the TT wall, since the V$_{\mathrm{O}}$ concentration is very large: the bound charge at the wall is not enough to compensate the positive charge and electrons migrate to the conduction band. Therefore, we adopt the 2x2x8 supercell described in the main text to show the influence of vacancies on the full system. In figure \ref{fig:pdosall} we show the density of states per TiO$_2$ plane with two V$_{\mathrm{O}}$ at the TT wall plane: there are HH wall states below the Fermi level as in the vacancy-free case, ensuring its $n$ type conductivity and showing the little influence of vacancies on its conducting properties. The unoccupied energy levels below the fermi level at the TT wall of the vacancy-free case have, nevertheless, disappeared, quenching the conductivity of the TT wall.

\providecommand{\noopsort}[1]{}\providecommand{\singleletter}[1]{#1}%

\end{document}